\documentclass[10pt, twocolumn, article]{revtex4-1}

\usepackage{graphicx}

\usepackage{amsmath}
\usepackage{amssymb}

\input epsf
\input rotate

\def\bea{\begin{eqnarray}}
\def\eea{\end{eqnarray}}
\def\ben{\begin{equation}}
\def\een{\end{equation}}
\def\benu{\begin{enumerate}}
\def\enu{\end{enumerate}}

\def\lsim {\ifmmode {\buildrel<\over\sim}}

\def\1var{(\bx_1...\bx\N)}

\def\half{\frac{1}{2}}

\def\br{{\bf r}}

\def\b1{{\bf 1}}
\def\bx{{x}}

\def\N{_{\sss N}}

\def\I{^{\rm I}}

\def\sph_int{ {\int d^3 r}}

\def\infintd3r{ \int_{-\infty}^\infty d^3r\,}
\def\intd3r{ \int d^3r\,}

\def\laplace1d{\frac{d^2}{dx^2}}
\def\plaplace1d{\frac{d^2}{d{x'}^2}}

\def\padr2{\frac{\partial^2}{\partial r^2}}

\def\N{{\cal N}}

\def\b{{\beta}}

\def\thesection{\arabic{section}}

\begin{document}

\title{{Resonance Lifetimes from Complex Densities}}
\author{Daniel L. Whitenack}
\affiliation{Department of Physics, Purdue
University, West Lafayette, IN 47907, USA}
\author{Adam Wasserman}
\affiliation{Department of Chemistry, Purdue
University, West Lafayette, IN 47907, USA}

\begin{abstract}

The {\em ab-initio} calculation of resonance lifetimes of metastable anions challenges modern quantum-chemical methods. The exact lifetime of the lowest-energy resonance is  encoded into a complex ``density" that can be obtained via complex-coordinate scaling. We illustrate this with one-electron examples and show how the lifetime can be extracted from the complex density in much the same way as the ground-state energy of bound systems is extracted from its ground-state density.

\end{abstract}

\maketitle

\section*{Introduction}

Shape and Feshbach resonances have been observed in many molecules both in the gas-phase and on surfaces.  These states often result in the formation of molecular negative ions whose energies and lifetimes can be measured experimentally~\cite{palmer}. Due to their diffuse electron clouds, finite lifetimes, and subtle manifestations of electron correlation effects, the calculation of the electronic structure of such metastable anions  poses special difficulties for the theorist (see recent review by Simons \cite{S08}). Obtaining accurate resonance lifetimes from first principles is a formidable 
task both because of the many-body nature of the problem (in particular bound-free correlation \cite{N00}) and because these states occur in the energy continuum where no normalizable wavefunction exists.

Standard Density Functional Theory (DFT) \cite{hohenberg,kohn} cannot come to the rescue. Even the {\em exact} exchange-correlation (XC) functional of DFT does not predict the true values for the anions' negative electron affinities because the correct solution converges in the infinite basis-set limit to the ground state of the corresponding neutral species (plus one electron off to infinity), and not to the metastable state of interest. Artificially binding the runaway electron with the use of a finite set of localized basis functions yields reasonable (approximate) resonance energies \cite{RTSN02}. Extrapolation techniques also work well for this purpose \cite{PVP08}, but the lifetimes are simply not accessible via standard DFT.  

Time-dependent DFT (TDDFT) \cite{RG84} is needed as a matter of principle for high-lying resonances (e.g. see ref.~\cite{KM09}  for autoionizing resonances), but  a complex-scaled version of {\em ground-state} DFT may be enough for the lowest-energy one. After all, in many respects this resonance plays the role of the `ground state' of the metastable anion \cite{WM07}.

This letter reports our proof-of-principle demonstration 
that the `ground-state' resonance lifetime is encoded in the complex resonance density, and that the lifetime can be extracted from this complex density  in much the same way as the ground-state energy is extracted in bound systems from the ground-state density. 

{\underline {\em Complex-Scaling:}}
For few-electron systems or simple model systems, the complex-scaling method 
\cite{moiseyev1,reinhardt,simon} is well-established and has drawn much attention since the original theorem of Balslev and Combes~\cite{balslev}.  
It has been applied successfully to study resonance phenomena in many branches of science: atomic collision processes \cite{H83}, acoustic resonances in tunnels \cite{HK08}, quantum-confined Stark effect in quantum wells \cite{DP05}, and many others. The method has been formulated to treat the many-electron problem within mean-field theories like Hartree-Fock \cite{M80} as well as straightforward configuration-interaction expansions \cite{J82}; it has also been used in combination with Density Functional Theory to study the broadening of levels in the vicinity of metal surfaces \cite{NT90,NT89,NT88}, or even with TDDFT to study above-threshold ionization of small negative ions \cite{TC02}. The fundamental question of scaling of density functionals upon coordinate-rotation in the complex plane, however, has not been investigated.

Here are the most relevant aspects of the complex-scaling technique \cite{reinhardt,M98}: When all coordinates of a many-electron Hamiltonian $\hat{H}$ are multiplied by a phase factor $e^{i\theta}$, the resulting non-hermitian operator $\hat{H}_\theta$ has the following properties for a suitable range of $\theta$ : (1) a discrete spectrum of real eigenvalues that is identical to that of the original (unrotated) Hamiltonian; (2) Scattering thresholds that are also identical to those of the original Hamiltonian; (3) continua beginning at each scattering threshold, rotated by an angle $2\theta$ with respect to those of the original Hamiltonian; and (4) Complex discrete eigenvalues appearing in the lower-half of the complex energy plane. Their corresponding eigenfuctions are square-integrable and can be associated with resonances whose energies and lifetimes are determined respectively by the real and imaginary parts of the eigenvalues.

\section*{Complex Densities}

We focus attention on the complex density $n_\theta(\br)$ associated with the lowest-energy resonance,
\ben
n_\theta(\br)=\langle \Psi_\theta^L|\hat{n}(\br)|\Psi_\theta^R\rangle
\label{eq:den}
\een
where $\hat{n}(\br)$ is the density operator, and $\langle\Psi_\theta^L|$ and $|\Psi_\theta^R\rangle$ are the left and right eigenvectors of $\hat{H}_\theta$ corresponding to the lowest-energy resonance. We require that $n_\theta(\br)$ be normalized to the number of electrons, as real densities are:
\ben
\int d\br n_\theta(\br)=N~~,
\label{e:norm}
\een
and ask whether $n_\theta(\br)$ contains the information about the {\em lifetime} of the resonance.  Our hypothesis, prompted by partial empirical evidence \cite{RTSN02} and previous observations \cite{E05,WM07}, is that the energy and lifetime of this resonance are determined respectively by the real and imaginary parts of the {\em ground-state} energy functional $E[n]$ evaluated at $n_\theta(\br)$. More specifically, the complex $E[n_\theta]$ is equal to:
\ben
{\cal{E}}[n_\theta]-\frac{i}{2}{\cal{L}}^{-1}[n_\theta]=F_\theta[n_\theta]+\int d\br n_\theta(\br) \textup{v}(\br e^{i\theta})
\label{e:hypothesis}
\een
where $\cal{E}$ is the lowest resonance energy, $\cal{L}$ the corresponding lifetime, $\textup{v}$ the external potential, and $F_\theta$ the complex-scaled Hohenberg-Kohn universal component of the energy density functional \cite{hohenberg}.

Our purpose here is to test the validity of Eq.(\ref{e:hypothesis}) for one-electron systems, where $F_\theta[n]$ is known exactly; it is given by $F_\theta[n]=e^{-2i\theta}T_W[n]$, where $T_W[n]$ is the Von Weizsacker functional \cite{weizsacker},
\ben
T_W[n]=\frac{1}{8}\int d\br \frac{|\nabla n(\br)|^2}{n(\br)}
\label{e:TW}
\een

{\em {\underline {Analytical Test:}}}
First, consider a radially-symmetric external potential given by $\textup{v}(r)=-\half r^{-2}+r^{-1}$. This is admittedly not too chemically meaningful but has the nice features that its resonances are known analytically \cite{D78}. It has no bound states but a resonance of energy ${\cal E}=1/4$ and lifetime ${\cal L}=2/\sqrt{3}$. Its complex-scaled Hamiltonian
\ben
\hat{H}_\theta=e^{-2i\theta}\left[\half\frac{d^2}{dr^2}+\frac{1}{r^2}-\frac{e^{i\theta}}{r}\right]
\label{e:H_analytic}
\een 
has the following radial eigenstate:
\ben
\phi_\theta(r)=N_\theta r^{(1-i\sqrt{3})/2}e^{i r k\theta}~~;~~k_\theta = -i e^{i\theta}(1+i\sqrt{3})/2
\label{e:eigenstate}
\een
The function $\phi_\theta(r)$ is not square-integrable for arbitrary $\theta$, but in the range $\left[\pi /6 < \theta < 7\pi /6 \right]$ it is, and the constant $N_\theta$ can be chosen in this range as $N_\theta = \left[(-2ik_\theta)^{2-i\sqrt{3}}/\Gamma(2-i\sqrt{3})\right]^{1/2}$, so the normalization condition for the resonance density, Eq.(\ref{e:norm}), is satisfied. Since the adjoint of the hamiltonian (\ref{e:H_analytic}) is simply equal to its complex-conjugate, the radial resonance density is given by $\tilde{n}_\theta (r)=N_\theta^2 r^{1-i\sqrt{3}} e^{2ik_\theta r}$ (we use a tilde on top of $n$ to denote it is the radial density, $\tilde{n}(r)=4\pi r^2 n(\br)$). Even though this function lies outside the domain of real densities for which the functional $F[n]$ was established \cite{hohenberg}, we plug it nevertheless into the right-hand-side of Eq.(\ref{e:hypothesis}) with $F_\theta[n_\theta]=e^{-2i\theta} T_W[n_\theta]$, and find the result: $E[n_\theta]=\frac{1}{4}-i\frac{\sqrt{3}}{4}$, whose real and (minus 2 times the) imaginary parts correspond respectively to the {\em correct} energy and inverse-lifetime of the metastable state. 

We note that the lifetime can also be extracted directly from the resonance density by looking at its asymptotic behavior ($r\to\infty$), just as the ionization energy $I$ of bound systems is extracted from the $r\to\infty$ decay of (real) ground-state densities. For one-electron systems, this decay is simply determined by the wavenumber $k=\sqrt{2I}$. Upon complex scaling, the `ionization energy' and `decay constant' both become complex ($I\to I_\theta$, and $k\to k_\theta$), and $I_\theta = \half e^{-2i\theta} k_\theta^2$. For the example below, using Eq.(\ref{e:eigenstate}), we find $I_\theta = -1/4+i\sqrt{3}/4$, correctly.

Next, a numerical example will allow us to look deeper into the meaning of the complex density, and how the resonance lifetime is extracted from it. 

\begin{figure}
  \scalebox{0.4}{\includegraphics{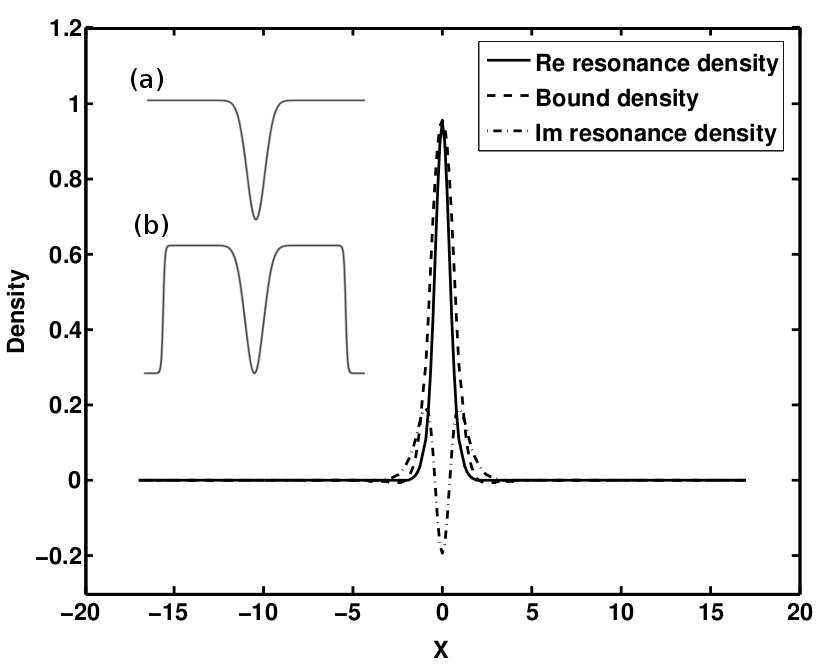}}
  \caption{The bound electron density of $\textup{v}_b(x)$ and the real and imaginary parts of the resonance density of $\textup{v}_u(x)$.  For these calculations $a=4$, $b=0.05$, $c=4$ and $d=10$.  Inset (a) shows the potential $\textup{v}_b(x)$ and inset (b) shows the potential $\textup{v}_u(x)$}
  \label{fgr:den}
\end{figure}

{\em {\underline{Numerical Test:}}}
We study the fate of the bound state of a simple one-dimensional gaussian well, $\textup{v}_b(x)= a(1 - e^{-\frac{x^2}{b}})$, when
%
%
steps are imposed on 
it at $\pm d$, making the problem unbound (see inset of \ref{fgr:den}), $\textup{v}_u(x)$:
\begin{equation}
 \textup{v}_u(x) = a\left( \frac{1}{1+e^{-2 c (x + d)}} - \frac{1}{1+e^{-2 c (x - d)}} - e^{-\frac{x^2}{b}} \right)
\end{equation}
\noindent
We use complex-scaling to solve for the complex eigenvalues and eigenfunctions.  As the steps are taken far away ($d\to\infty$),   the resonance energy and the real part of the resonance density approach the bound state energy and ground-state density of the bound potential $\textup{v}_b(x)$, respectively, as expected \cite{buchleitner, taylor} (\ref{fgr:den}).


For example, when $a = 4$ and $b = 0.05$ the bound-state energy of an electron in $\textup{v}_b(x)$ is 1.6246.  
\ref{tab:reseng} shows that this same value is obtained for $\textup{v}_u(x)$ when $d\geq 2$.  
Also, the imaginary part of the resonance energy shrinks exponentially as the steps get wider.  This behavior is expected because the inverse of $\Im E_\theta$ corresponds to the lifetime of the resonance, and the resonance is long lived for wide steps in $\textup{v}_u(x)$.

\begin{table}
\caption{Resonance energies for $\textup{v}_u (x)$ with various values of $d$.  For all calculations $c=4$, $a=4$, $b=0.05$ 
.}
\begin{center}
\begin{tabular}{lclcl}
$d$ & \quad $\Re E_\theta$ & \quad \quad \quad $\Im E_\theta $ \\
\hline
2 & \quad $1.6285$ & \quad $-3.3630 \times 10^{-3}$ \\
3 & \quad $1.6246$ & \quad $-4.2189 \times 10^{-5}$ \\
5 & \quad $1.6246$ & \quad $-1.3768 \times 10^{-9}$ \\
10 & \quad $1.6246$ & \quad $-1.7484 \times 10^{-14}$ \\
\hline
\end{tabular}
\end{center}
\label{tab:reseng}
\end{table}

The Fourier Grid Hamiltonian technique (FGH) \cite{clay,chu} is found to be the most efficient method for these complex scaling calculations.  It allows one to solve the eigenvalue problem without the use of a basis set expansion.  The Hamiltonian is written in coordinate representation, $\langle \vec{x}|\hat{H}_\theta|\vec{x}' \rangle$, and the spatial coordinate is discretized.  This matrix is then diagonlized and the resulting eigenvectors give directly the amplitude of the solutions of the resonance wave function.  Using this method, the resonance energy for $\textup{v}_b(x)$ is converged using less than 100 grid points. 

\begin{figure}
  \scalebox{0.3}{\includegraphics{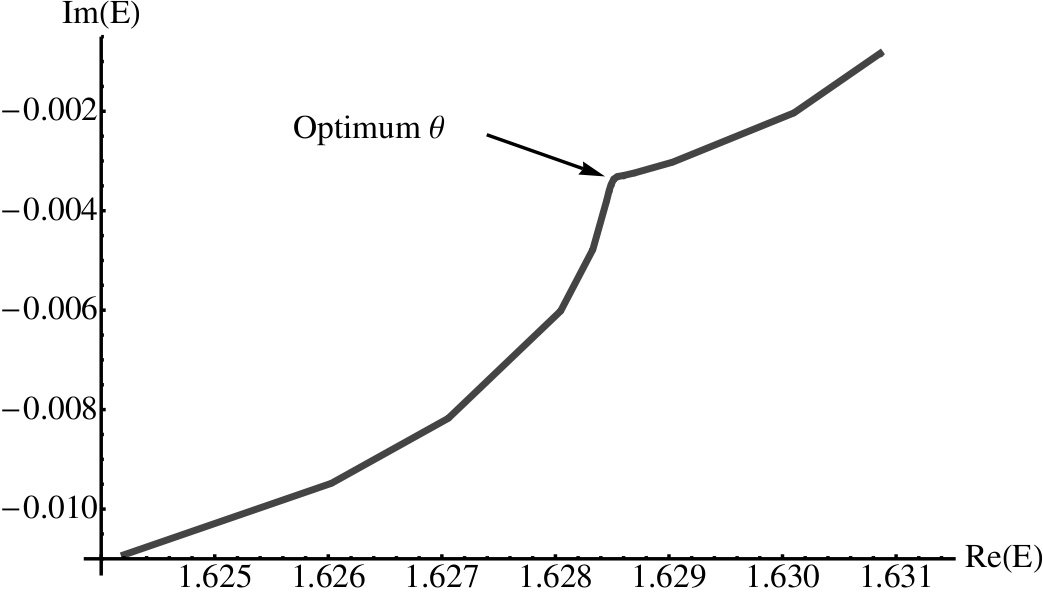}}
  \caption{An example $\theta$ trajectory.  In this case the real part of the energy is plotted against the imaginary part of the energy for $\theta$ values from 0.24 to 0.54, and the optimum $\theta$ is found at the kink with a value of 0.312.}
  \label{fgr:thetatraj}
\end{figure}

The resonance energy and lifetime is independent of $\theta$ \cite{moiseyev1}.  However when doing a numerical calculation, the use of finite basis sets or finite grids necessitates finding an optimum $\theta$.  This optimum condition is acheived by examining $\theta$-trajectories \cite{moiseyev1} (see \ref{fgr:thetatraj}).  When the real and imaginary part of the resonance energy become stationary around the optimum value of $\theta$ one can see a kink or loop in the trajectory. Within our range of parameters, the optimum $\theta$ for $\textup{v}_u(x)$ falls between $0.28$ and $0.4$.  With many more grid points this procedure becomes less important as all the theta trajectory points collapse to the region of the resonance.

\begin{figure}
  \scalebox{0.4}{\includegraphics{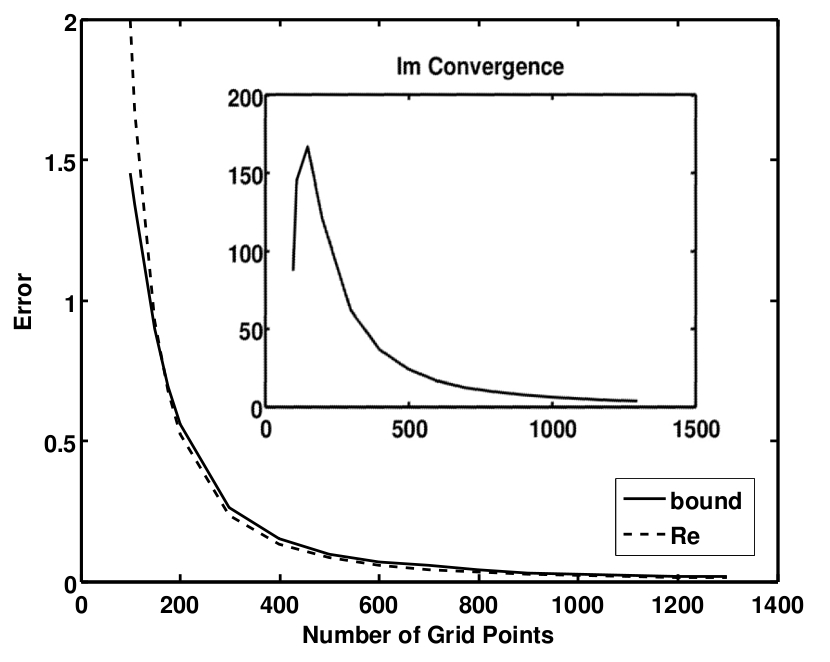}}
  \caption{The \% error in the real and imaginary part of the resonance energy calculated from Eq.\ref{e:hypothesis} with $F_\theta[n_\theta]=e^{-2i\theta} T_W[n_\theta]$ as a function of the number of grid point used in the calculation.  For this plot $a=4$, $b=0.05$, $c=4$ and $d=2$.  Also included is the convergence of the bound energy of $\textup{v}_b(x)$ calculated from the unscaled $T_W[n]$ functional.}
  \label{fgr:conv}
\end{figure}

The key point is this: When $n_\theta(x)$ is normalized according to Eq.~\ref{e:norm}, the correct resonance energies and lifetimes are indeed obtained from $e^{-2i\theta}T_W[n_\theta]$. For example, with $a=4$, $b=0.05$, $c=4$ and $d=2$ the resonance energy is $1.6285 - 3.3630 \times 10^{-3} i$.  Using the FGH method with 999 grid points, the von Weizsacker functional gives a resonance energy of $1.6282 - 3.1126 \times 10^{-3}i$.  The convergence of the functional calculations is included in \ref{fgr:conv}, where the error is defined relative to FGH results obtained using the same number of grid points. The error of the resonance energies obtained with the functional is less than 1\% for more than 200 grid points, and the error of the lifetimes is less than 1\% for more than 2500 grid points.  
%
%
From \ref{fgr:conv}, it is clear that the real part of the resonance density converges very quickly and its behavior is comparable to that of the bound energy convergence obtained for $\textup{v}_b(x)$ via $T_W[n]$.  The convergence of the imaginary part follows a similar trend, but errors of less than $10\%$ are only obtained when more than 1000 grid points are employed. The imaginary functional is {\em exact}, but sensitive to small changes in the resonance density.  


\section*{Conclusion}

For one-electron systems, we have shown that the lowest-energy resonance lifetime is encoded into the corresponding complex resonance density, and can be extracted from it via the properly scaled density functional. If the same result was established for many-electron systems, a new way of calculating `ground-state' resonance lifetimes could be based on a complex-scaled version of standard KS-DFT, an exciting prospect. We are working along these lines. 



\begin{acknowledgements}

Support from the Petroleum Research Fund grant No.PRF\# 49599-DNI6, and discussions with Yu Zhang, are gratefully acknowledged.
\end{acknowledgements}

\bibliography{ksres1}

\begin{thebibliography}{10}%
\makeatletter
\providecommand \@ifxundefined [1]{%
 \ifx #1\undefined \expandafter \@firstoftwo
 \else \expandafter \@secondoftwo
\fi
}%
\providecommand \@ifnum [1]{%
 \ifnum #1\expandafter \@firstoftwo
 \else \expandafter \@secondoftwo
\fi
}%
\providecommand \enquote [1]{``#1''}%
\providecommand \bibnamefont  [1]{#1}%
\providecommand \bibfnamefont [1]{#1}%
\providecommand \citenamefont [1]{#1}%
\providecommand\href[0]{\@sanitize\@href}%
\providecommand\@href[1]{\endgroup\@@startlink{#1}\endgroup\@@href}%
\providecommand\@@href[1]{#1\@@endlink}%
\providecommand \@sanitize [0]{\begingroup\catcode`\&12\catcode`\#12\relax}%
\@ifxundefined \pdfoutput {\@firstoftwo}{%
 \@ifnum{\z@=\pdfoutput}{\@firstoftwo}{\@secondoftwo}%
}{%
 \providecommand\@@startlink[1]{\leavevmode\special{html:<a href="#1">}}%
 \providecommand\@@endlink[0]{\special{html:</a>}}%
}{%
 \providecommand\@@startlink[1]{%
  \leavevmode
  \pdfstartlink
   attr{/Border[0 0 1 ]/H/I/C[0 1 1]}%
   user{/Subtype/Link/A<</Type/Action/S/URI/URI(#1)>>}%
  \relax
 }%
 \providecommand\@@endlink[0]{\pdfendlink}%
}%
\providecommand \url  [0]{\begingroup\@sanitize \@url }%
\providecommand \@url [1]{\endgroup\@href {#1}{\urlprefix}}%
\providecommand \urlprefix [0]{URL }%
\providecommand \Eprint[0]{\href }%
\@ifxundefined \urlstyle {%
  \providecommand \doi [1]{doi:\discretionary{}{}{}#1}%
}{%
  \providecommand \doi [0]{doi:\discretionary{}{}{}\begingroup
  \urlstyle{rm}\Url }%
}%
\providecommand \doibase [0]{http://dx.doi.org/}%
\providecommand \Doi[1]{\href{\doibase#1}}%
\providecommand \bibAnnote [3]{%
  \BibitemShut{#1}%
  \begin{quotation}\noindent
    \textsc{Key:}\ #2\\\textsc{Annotation:}\ #3%
  \end{quotation}%
}%
\providecommand \bibAnnoteFile [2]{%
  \IfFileExists{#2}{\bibAnnote {#1} {#2} {\input{#2}}}{}%
}%
\providecommand \typeout [0]{\immediate \write \m@ne }%
\providecommand \selectlanguage [0]{\@gobble}%
\providecommand \bibinfo [0]{\@secondoftwo}%
\providecommand \bibfield [0]{\@secondoftwo}%
\providecommand \translation [1]{[#1]}%
\providecommand \BibitemOpen[0]{}%
\providecommand \bibitemStop [0]{}%
\providecommand \bibitemNoStop [0]{.\EOS\space}%
\providecommand \EOS [0]{\spacefactor3000\relax}%
\providecommand \BibitemShut [1]{\csname bibitem#1\endcsname}%
\bibitem{palmer}%
  \BibitemOpen
  \bibfield{author}{%
  \bibinfo {author} {\bibfnamefont{R.~E.}\ \bibnamefont{Palmer}}\ and\ \bibinfo
  {author} {\bibfnamefont{P.~J.}\ \bibnamefont{Rous}},\ }%
  \bibfield{journal}{%
  \bibinfo {journal} {Rev. Mod. Phys.}\ }%
  \textbf{\bibinfo {volume} {64}},\ \bibinfo {pages} {383} (\bibinfo {year}
  {1992})%
  \bibAnnoteFile{NoStop}{palmer}%
\bibitem{S08}%
  \BibitemOpen
  \bibfield{author}{%
  \bibinfo {author} {\bibfnamefont{J.}~\bibnamefont{Simons}},\ }%
  \bibfield{journal}{%
  \bibinfo {journal} {J. Phys. Chem. A}\ }%
  \textbf{\bibinfo {volume} {112}},\ \bibinfo {pages} {6401} (\bibinfo {year}
  {2008})%
  \bibAnnoteFile{NoStop}{S08}%
\bibitem{N00}%
  \BibitemOpen
  \bibfield{author}{%
  \bibinfo {author} {\bibfnamefont{R.~K.}\ \bibnamefont{Nesbet}},\ }%
  \bibfield{journal}{%
  \bibinfo {journal} {Phys. Rev. A}\ }%
  \textbf{\bibinfo {volume} {62}},\ \bibinfo {pages} {040701(R)} (\bibinfo
  {year} {2000})%
  \bibAnnoteFile{NoStop}{N00}%
\bibitem{hohenberg}%
  \BibitemOpen
  \bibfield{author}{%
  \bibinfo {author} {\bibfnamefont{P.}~\bibnamefont{Hohenberg}}\ and\ \bibinfo
  {author} {\bibfnamefont{W.}~\bibnamefont{Kohn}},\ }%
  \bibfield{journal}{%
  \bibinfo {journal} {Phys. Rev.}\ }%
  \textbf{\bibinfo {volume} {136}},\ \bibinfo {pages} {B864} (\bibinfo {year}
  {1964})%
  \bibAnnoteFile{NoStop}{hohenberg}%
\bibitem{kohn}%
  \BibitemOpen
  \bibfield{author}{%
  \bibinfo {author} {\bibfnamefont{W.}~\bibnamefont{Kohn}}\ and\ \bibinfo
  {author} {\bibfnamefont{L.~J.}\ \bibnamefont{Sham}},\ }%
  \bibfield{journal}{%
  \bibinfo {journal} {Phys. Rev.}\ }%
  \textbf{\bibinfo {volume} {140}},\ \bibinfo {pages} {A1133} (\bibinfo {year}
  {1965})%
  \bibAnnoteFile{NoStop}{kohn}%
\bibitem{RTSN02}%
  \BibitemOpen
  \bibfield{author}{%
  \bibinfo {author} {\bibfnamefont{J.~C.}\ \bibnamefont{Rienstra-Kiracofe}},
  \bibinfo {author} {\bibfnamefont{G.~S.}\ \bibnamefont{Tschumper}}, \bibinfo
  {author} {\bibfnamefont{H.~F.}\ \bibnamefont{Schaefer}}, \bibinfo {author}
  {\bibfnamefont{S.}~\bibnamefont{Nandi}},\ and\ \bibinfo {author}
  {\bibfnamefont{G.~B.}\ \bibnamefont{Elliso}},\ }%
  \bibfield{journal}{%
  \bibinfo {journal} {Chem. Rev.}\ }%
  \textbf{\bibinfo {volume} {102}},\ \bibinfo {pages} {231} (\bibinfo {year}
  {2002})%
  \bibAnnoteFile{NoStop}{RTSN02}%
\bibitem{PVP08}%
  \BibitemOpen
  \bibfield{author}{%
  \bibinfo {author} {\bibfnamefont{M.}~\bibnamefont{Puiatti}}, \bibinfo
  {author} {\bibfnamefont{D.~M.~A.}\ \bibnamefont{Vera}},\ and\ \bibinfo
  {author} {\bibfnamefont{A.~B.}\ \bibnamefont{Pierini}},\ }%
  \bibfield{journal}{%
  \bibinfo {journal} {Phys. Chem. Chem. Phys.}\ }%
  \textbf{\bibinfo {volume} {10}},\ \bibinfo {pages} {1394} (\bibinfo {year}
  {2008})%
  \bibAnnoteFile{NoStop}{PVP08}%
\bibitem{RG84}%
  \BibitemOpen
  \bibfield{author}{%
  \bibinfo {author} {\bibfnamefont{E.}~\bibnamefont{Runge}}\ and\ \bibinfo
  {author} {\bibfnamefont{E.~K.~U.}\ \bibnamefont{Gross}},\ }%
  \bibfield{journal}{%
  \bibinfo {journal} {Phys. Rev. Lett.}\ }%
  \textbf{\bibinfo {volume} {52}},\ \bibinfo {pages} {997} (\bibinfo {year}
  {1984})%
  \bibAnnoteFile{NoStop}{RG84}%
\bibitem{KM09}%
  \BibitemOpen
  \bibfield{author}{%
  \bibinfo {author} {\bibfnamefont{A.~J.}\ \bibnamefont{Krueger}}\ and\
  \bibinfo {author} {\bibfnamefont{N.~T.}\ \bibnamefont{Maitra}},\ }%
  \bibfield{journal}{%
  \bibinfo {journal} {Phys. Chem. Chem. Phys.}\ }%
  \textbf{\bibinfo {volume} {11}},\ \bibinfo {pages} {4655} (\bibinfo {year}
  {2009})%
  \bibAnnoteFile{NoStop}{KM09}%
\bibitem{WM07}%
  \BibitemOpen
  \bibfield{author}{%
  \bibinfo {author} {\bibfnamefont{A.}~\bibnamefont{Wasserman}}\ and\ \bibinfo
  {author} {\bibfnamefont{N.}~\bibnamefont{Moiseyev}},\ }%
  \bibfield{journal}{%
  \bibinfo {journal} {Phys. Rev. Lett.}\ }%
  \textbf{\bibinfo {volume} {98}},\ \bibinfo {pages} {093003} (\bibinfo {year}
  {2007})%
  \bibAnnoteFile{NoStop}{WM07}%
\bibitem{moiseyev1}%
  \BibitemOpen
  \bibfield{author}{%
  \bibinfo {author} {\bibfnamefont{N.}~\bibnamefont{Moiseyev}}, \bibinfo
  {author} {\bibfnamefont{P.~R.}\ \bibnamefont{Certain}},\ and\ \bibinfo
  {author} {\bibfnamefont{F.}~\bibnamefont{Weinhold}},\ }%
  \bibfield{journal}{%
  \bibinfo {journal} {Mol. Phys.}\ }%
  \textbf{\bibinfo {volume} {36}},\ \bibinfo {pages} {1613} (\bibinfo {year}
  {1978})%
  \bibAnnoteFile{NoStop}{moiseyev1}%
\bibitem{reinhardt}%
  \BibitemOpen
  \bibfield{author}{%
  \bibinfo {author} {\bibfnamefont{W.~P.}\ \bibnamefont{Reinhardt}},\ }%
  \bibfield{journal}{%
  \bibinfo {journal} {Ann. Rev. Phys. Chem.}\ }%
  \textbf{\bibinfo {volume} {33}},\ \bibinfo {pages} {223} (\bibinfo {year}
  {1982})%
  \bibAnnoteFile{NoStop}{reinhardt}%
\bibitem{simon}%
  \BibitemOpen
  \bibfield{author}{%
  \bibinfo {author} {\bibfnamefont{B.}~\bibnamefont{Simon}},\ }%
  \bibfield{journal}{%
  \bibinfo {journal} {Ann. Math.}\ }%
  \textbf{\bibinfo {volume} {97}},\ \bibinfo {pages} {247} (\bibinfo {year}
  {1973})%
  \bibAnnoteFile{NoStop}{simon}%
\bibitem{balslev}%
  \BibitemOpen
  \bibfield{author}{%
  \bibinfo {author} {\bibfnamefont{E.}~\bibnamefont{Balslev}}\ and\ \bibinfo
  {author} {\bibfnamefont{J.~M.}\ \bibnamefont{Combes}},\ }%
  \bibfield{journal}{%
  \bibinfo {journal} {Commun. Math. Phys.}\ }%
  \textbf{\bibinfo {volume} {22}},\ \bibinfo {pages} {280} (\bibinfo {year}
  {1971})%
  \bibAnnoteFile{NoStop}{balslev}%
\bibitem{H83}%
  \BibitemOpen
  \bibfield{author}{%
  \bibinfo {author} {\bibfnamefont{Y.~K.}\ \bibnamefont{Ho}},\ }%
  \bibfield{journal}{%
  \bibinfo {journal} {Phys. Lett.}\ }%
  \textbf{\bibinfo {volume} {99}},\ \bibinfo {pages} {1} (\bibinfo {year}
  {1983})%
  \bibAnnoteFile{NoStop}{H83}%
\bibitem{HK08}%
  \BibitemOpen
  \bibfield{author}{%
  \bibinfo {author} {\bibfnamefont{S.}~\bibnamefont{Hein}}\ and\ \bibinfo
  {author} {\bibfnamefont{W.}~\bibnamefont{Koch}},\ }%
  \bibfield{journal}{%
  \bibinfo {journal} {J. Fluid. Mech.}\ }%
  \textbf{\bibinfo {volume} {605}},\ \bibinfo {pages} {401} (\bibinfo {year}
  {2008})%
  \bibAnnoteFile{NoStop}{HK08}%
\bibitem{DP05}%
  \BibitemOpen
  \bibfield{author}{%
  \bibinfo {author} {\bibfnamefont{L.~I.}\ \bibnamefont{Deych}}\ and\ \bibinfo
  {author} {\bibfnamefont{I.~V.}\ \bibnamefont{Pnonomarev}},\ }%
  \bibfield{journal}{%
  \bibinfo {journal} {Phys. Rev. B}\ }%
  \textbf{\bibinfo {volume} {71}},\ \bibinfo {pages} {035342} (\bibinfo {year}
  {2005})%
  \bibAnnoteFile{NoStop}{DP05}%
\bibitem{M80}%
  \BibitemOpen
  \bibfield{author}{%
  \bibinfo {author} {\bibfnamefont{C.~W.}\ \bibnamefont{McCurdy}},\ }%
  \bibfield{journal}{%
  \bibinfo {journal} {Phys. Rev. A}\ }%
  \textbf{\bibinfo {volume} {21}},\ \bibinfo {pages} {464} (\bibinfo {year}
  {1980})%
  \bibAnnoteFile{NoStop}{M80}%
\bibitem{J82}%
  \BibitemOpen
  \bibfield{author}{%
  \bibinfo {author} {\bibfnamefont{B.~R.}\ \bibnamefont{Junker}},\ }%
  \bibfield{journal}{%
  \bibinfo {journal} {Adv. At. Mol. Phys.}\ }%
  \textbf{\bibinfo {volume} {18}},\ \bibinfo {pages} {207} (\bibinfo {year}
  {1982})%
  \bibAnnoteFile{NoStop}{J82}%
\bibitem{NT90}%
  \BibitemOpen
  \bibfield{author}{%
  \bibinfo {author} {\bibfnamefont{P.}~\bibnamefont{Nordlander}}\ and\ \bibinfo
  {author} {\bibfnamefont{J.~C.}\ \bibnamefont{Tully}},\ }%
  \bibfield{journal}{%
  \bibinfo {journal} {Phys. Rev. B}\ }%
  \textbf{\bibinfo {volume} {42}},\ \bibinfo {pages} {5564} (\bibinfo {year}
  {1990})%
  \bibAnnoteFile{NoStop}{NT90}%
\bibitem{NT89}%
  \BibitemOpen
  \bibfield{author}{%
  \bibinfo {author} {\bibfnamefont{P.}~\bibnamefont{Nordlander}}\ and\ \bibinfo
  {author} {\bibfnamefont{J.~C.}\ \bibnamefont{Tully}},\ }%
  \bibfield{journal}{%
  \bibinfo {journal} {Surf. Sci.}\ }%
  \textbf{\bibinfo {volume} {211}},\ \bibinfo {pages} {207} (\bibinfo {year}
  {1989})%
  \bibAnnoteFile{NoStop}{NT89}%
\bibitem{NT88}%
  \BibitemOpen
  \bibfield{author}{%
  \bibinfo {author} {\bibfnamefont{P.}~\bibnamefont{Nordlander}}\ and\ \bibinfo
  {author} {\bibfnamefont{J.~C.}\ \bibnamefont{Tully}},\ }%
  \bibfield{journal}{%
  \bibinfo {journal} {Phys. Rev. Lett.}\ }%
  \textbf{\bibinfo {volume} {61}},\ \bibinfo {pages} {990} (\bibinfo {year}
  {1988})%
  \bibAnnoteFile{NoStop}{NT88}%
\bibitem{TC02}%
  \BibitemOpen
  \bibfield{author}{%
  \bibinfo {author} {\bibfnamefont{D.~A.}\ \bibnamefont{Telnov}}\ and\ \bibinfo
  {author} {\bibfnamefont{S.~I.}\ \bibnamefont{Chu}},\ }%
  \bibfield{journal}{%
  \bibinfo {journal} {Phys. Rev. A}\ }%
  \textbf{\bibinfo {volume} {66}},\ \bibinfo {pages} {043417} (\bibinfo {year}
  {2002})%
  \bibAnnoteFile{NoStop}{TC02}%
\bibitem{M98}%
  \BibitemOpen
  \bibfield{author}{%
  \bibinfo {author} {\bibfnamefont{N.}~\bibnamefont{Moiseyev}},\ }%
  \bibfield{journal}{%
  \bibinfo {journal} {Phys. Lett.}\ }%
  \textbf{\bibinfo {volume} {33}},\ \bibinfo {pages} {1982} (\bibinfo {year}
  {1982})%
  \bibAnnoteFile{NoStop}{M98}%
\bibitem{E05}%
  \BibitemOpen
  \bibfield{author}{%
  \bibinfo {author} {\bibfnamefont{M.}~\bibnamefont{Ernzerhof}},\ }%
  \bibfield{journal}{%
  \bibinfo {journal} {J. Chem. Phys.}\ }%
  \textbf{\bibinfo {volume} {125}},\ \bibinfo {pages} {124104} (\bibinfo {year}
  {2006})%
  \bibAnnoteFile{NoStop}{E05}%
\bibitem{weizsacker}%
  \BibitemOpen
  \bibfield{author}{%
  \bibinfo {author} {\bibfnamefont{C.~F.}\ \bibnamefont{von Weizsacker}},\ }%
  \bibfield{journal}{%
  \bibinfo {journal} {Z. f. Phys.}\ }%
  \textbf{\bibinfo {volume} {96}},\ \bibinfo {pages} {431} (\bibinfo {year}
  {1935})%
  \bibAnnoteFile{NoStop}{weizsacker}%
\bibitem{D78}%
  \BibitemOpen
  \bibfield{author}{%
  \bibinfo {author} {\bibfnamefont{G.~D.}\ \bibnamefont{Doolen}},\ }%
  \bibfield{journal}{%
  \bibinfo {journal} {Int. J. Quant. Chem.}\ }%
  \textbf{\bibinfo {volume} {14}},\ \bibinfo {pages} {523} (\bibinfo {year}
  {1978})%
  \bibAnnoteFile{NoStop}{D78}%
\bibitem{buchleitner}%
  \BibitemOpen
  \bibfield{author}{%
  \bibinfo {author} {\bibfnamefont{A.}~\bibnamefont{Buchleitner}}, \bibinfo
  {author} {\bibfnamefont{B.}~\bibnamefont{Gr\'{e}maud}},\ and\ \bibinfo
  {author} {\bibfnamefont{D.}~\bibnamefont{Delande}},\ }%
  \bibfield{journal}{%
  \bibinfo {journal} {J. Phys. B: At. Mol. Phys.}\ }%
  \textbf{\bibinfo {volume} {27}},\ \bibinfo {pages} {2663} (\bibinfo {year}
  {1994})%
  \bibAnnoteFile{NoStop}{buchleitner}%
\bibitem{taylor}%
  \BibitemOpen
  \bibfield{author}{%
  \bibinfo {author} {\bibfnamefont{J.~R.}\ \bibnamefont{Taylor}},\ }%
  \emph{\bibinfo {title} {Scattering Theory}}\ (\bibinfo {publisher} {Dover
  Publications, Inc.},\ \bibinfo {address} {New York},\ \bibinfo {year}
  {1972})%
  \bibAnnoteFile{NoStop}{taylor}%
\bibitem{clay}%
  \BibitemOpen
  \bibfield{author}{%
  \bibinfo {author} {\bibfnamefont{C.~C.}\ \bibnamefont{Marston}}\ and\
  \bibinfo {author} {\bibfnamefont{G.~G.}\ \bibnamefont{Balint-Kurti}},\ }%
  \bibfield{journal}{%
  \bibinfo {journal} {J. Chem. Phys.}\ }%
  \textbf{\bibinfo {volume} {91}},\ \bibinfo {pages} {3571} (\bibinfo {year}
  {1989})%
  \bibAnnoteFile{NoStop}{clay}%
\bibitem{chu}%
  \BibitemOpen
  \bibfield{author}{%
  \bibinfo {author} {\bibfnamefont{S.}~\bibnamefont{Chu}},\ }%
  \bibfield{journal}{%
  \bibinfo {journal} {Chem. Phys. Lett.}\ }%
  \textbf{\bibinfo {volume} {167}},\ \bibinfo {pages} {155} (\bibinfo {year}
  {1990})%
  \bibAnnoteFile{NoStop}{chu}%
\end{thebibliography}%

\end{document}